\documentclass[aps,prl,twocolumn,showpacs,superscriptaddress]{revtex4-1}  
\usepackage{graphicx}  
\usepackage{dcolumn}   
\usepackage{bm}        
\usepackage{amssymb}   

\begin{document}

\title{Charge order in LuFe$_{2}$O$_{4}$: an unlikely route to ferroelectricity}

\author{J. de Groot}
\affiliation{Peter Gr\"{u}nberg Institut PGI and J\"{u}lich Centre for Neutron Science JCNS, JARA-FIT, Forschungszentrum J\"{u}lich GmbH, 52425 J\"{u}lich, Germany}
\author{T. Mueller}
\affiliation{Peter Gr\"{u}nberg Institut PGI and J\"{u}lich Centre for Neutron Science JCNS, JARA-FIT, Forschungszentrum J\"{u}lich GmbH, 52425 J\"{u}lich, Germany}
\author{R.A. Rosenberg}
\affiliation{Argonne National Laboratory, 9700 S. Cass Avenue, Argonne, IL 60439, USA}
\author{D.J. Keavney}
\affiliation{Argonne National Laboratory, 9700 S. Cass Avenue, Argonne, IL 60439, USA}
\author{Z. Islam}
\affiliation{Argonne National Laboratory, 9700 S. Cass Avenue, Argonne, IL 60439, USA}
\author{J.-W. Kim}
\affiliation{Argonne National Laboratory, 9700 S. Cass Avenue, Argonne, IL 60439, USA}
\author{M. Angst}
\email{M.Angst@fz-juelich.de}
\affiliation{Peter Gr\"{u}nberg Institut PGI and J\"{u}lich Centre for Neutron Science JCNS, JARA-FIT, Forschungszentrum J\"{u}lich GmbH, 52425 J\"{u}lich, Germany}

\date{\today}

\begin{abstract}
We present the refinement of the crystal structure of charge-ordered LuFe$_{2}$O$_{4}$, based on single-crystal x-ray diffraction data. The arrangement of the different Fe-valence states, determined with bond-valence-sum analysis, corresponds to a stacking of charged Fe bilayers, in contrast to the polar bilayers previously suggested. This arrangement is supported by an analysis of x-ray magnetic circular dichroism spectra, which also evidences a strong charge-spin coupling. The non-polar bilayers are inconsistent with charge order based ferroelectricity.
\end{abstract}

\maketitle 
Multiferroic oxides with strong magnetoelectric coupling are of high interest for potential information-technology applications \cite{Wang2009a}, but they are rare because the traditional mechanism of ferroelectricity is incompatible with magnetism \cite{Hill2000}. Consequently, much attention is focused on various unconventional mechanisms of ferroelectricity \cite{Wang2009a}. Of these, ferroelectricity originating from charge ordering (CO) is particularly intriguing because it potentially combines large electric polarizations with strong magnetoelectric coupling \cite{vanBrink}. However, examples of oxides where this mechanism occurs are exceedingly rare and none is really well understood.\par 

\begin{figure}[b]
\includegraphics[width=0.99 \linewidth]{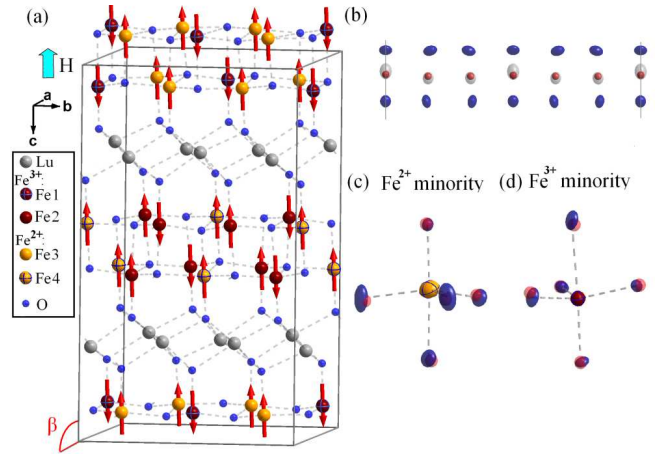}
\caption{\label{fig:Phaselow} (Color online) (a) Monoclinic crystal structure $C2/m$ of LuFe$_{2}$O$_{4}$ ($a\!=\!5.95\,$\AA, $b\!=\!10.30\,$\AA, $c\!=\!16.96\,$\AA$\,$, $\beta\!=\!96.72^{\circ}$). The refined data was measured at 210$\,$K. The ferrimagnetic high-field spin order and Fe$^{3+/2+}$ charge order is represented by arrows and different colors respectively. (b) Lu and O atoms drawn as thermal ellipsoids in projection along $a$. For comparison the Lu positions at 350$\,$K are displayed as spheres. (c,d) O coordination at 210$\,$K and 350$\,$K for Fe$^{2+/3+}$ minority (red spheres indicate the O-positions at 350$\,$K).}
\end{figure}

LuFe$_{2}$O$_{4}$ is often cited as the prototypical example of CO-based ferroelectricity (see, e.g., \cite{Wang2009a}). In this material, Fe valence order has been proposed to render the triangular Fe/O bilayers (see Fig.~\ref{fig:Phaselow}a) polar by making one of the two layers rich in Fe$^{2+}$ and the other in Fe$^{3+}$ \cite{Ikeda}. Because of this new mechanism of ferroelectricity and also because of the high transition temperatures of CO ($T_{CO}\!\sim$320$\,$K) and magnetism ($T_{N}\!\sim$240$\,$K) LuFe$_{2}$O$_{4}$ is increasingly attracting attention \cite{Nagano,Xiang,HARRIS,Mulders2,XMCD,WU,d0det,WEN09,Angst1,REN,XU,WEN10,ROUQUETTE,MULDERS,XMCD2,XU2,Michiuchi2009,Zeng2008,Li2008,JOOST,Fisher2011,REN}. That the Fe/O bilayers become polar upon CO has never been challenged. Symmetry-analysis of CO superstructure reflections \cite{Angst1} led to the proposal of an antiferroelectric stacking of the bilayer polarizations in the ground state, but did not cast into doubt the polar nature of the CO bilayers. Although these polar bilayers are generally accepted in the LuFe$_{2}$O$_{4}$ literature \cite{Mulders2,Nagano,Xiang,HARRIS,XMCD,WEN09}, a direct proof is lacking. An assumption-free experimental determination of whether or not the CO in the Fe/O bilayers is polar would be crucial given the dependence of the proposed mechanism of ferroelectricity in LuFe$_{2}$O$_{4}$ on polar bilayers.

In this letter, we present the first crystal structural refinement taking into account the superstructure due to CO in LuFe$_{2}$O$_{4}$, performed on single-crystal x-ray diffraction data. Identifying the positions of Fe$^{2+}$ and Fe$^{3+}$ valences in the structure with bond-valence-sum (BVS) analysis, an unexpected new CO pattern with charged Fe/O bilayers emerges (Fig.~\ref{fig:Phaselow}a). We also present x-ray magnetic circular dichroism (XMCD) measurements, which link (Fig.~\ref{fig:Phaselow}a) the CO with the spin order determined elsewhere \cite{JOOST}, further corroborating the new CO pattern. This new CO arrangement with charged and {\em non-polar} bilayers is in strong contrast to all previously suggested CO configurations with polar bilayers \cite{YAMADA2,Ikeda,Angst1}. We discuss the implications of this result on ``{\em ferroelectricity from CO}$\,$" in LuFe$_{2}$O$_{4}$, addressing the possibility of polarizing the bilayers by an electric field. Finally, we also address the relevance of the strict spin-charge coupling to the CO transition. 

\begin{figure}
\includegraphics[width=0.99 \linewidth]{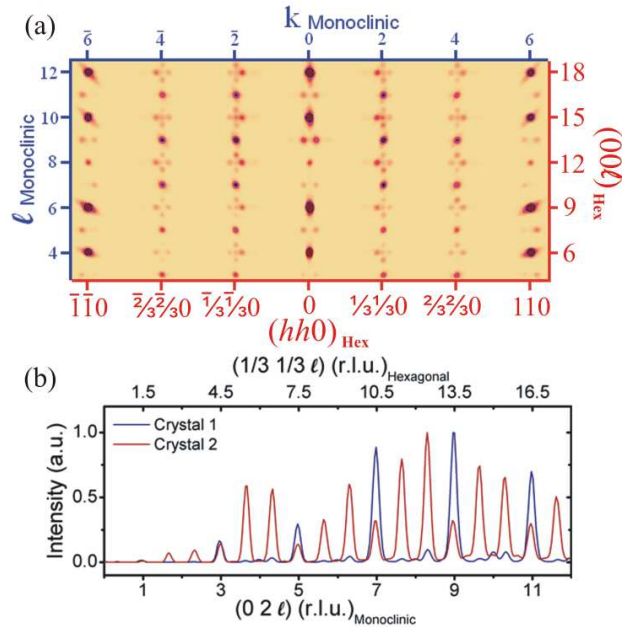}
\caption{\label{fig:ACMS} (Color online) (a) Composite precession image of crystal 1 in the $(0\,k\,\ell)_{\mathrm{Mon}}$-plane indexed in the new monoclinic cell, measured at 210$\,$K. (b) Intensity distribution along $(0\,2\,\ell)_{\mathrm{Mon}}$ for two crystals at 210$\,$K.}
\end{figure}

\begin{table}
\begin{center}
\caption{\label{tab:Valences} Valences from bond-valence-sum for different Fe-sites at 210K [$C2/m$] and 350K [$R\overline{3}m$].}
\begin{tabular}{ l c c c c}
\hline
\hline
Site & T[K] &  $\ $  $\left\langle (Fe-O) \right\rangle $$\!$ [{\AA}]  $\ $  &  $\ $  $V$ from BVS   $\ $ & Wyckoff  \\
\hline

Fe$_{R\overline{3}m}$& 350 & 2.030 & 2.38(3) & 6c \\
\hline

Fe1                  & 210 & 1.998 & 2.91(2) & 4i \\

Fe2                  & 210 & 1.999 & 2.75(2) & 8j \\

Fe3                  & 210 & 2.058 & 2.10(1) & 8j \\

Fe4                  & 210 & 2.100 & 1.92(1) & 4i \\
\hline
\hline
\end{tabular}
\end{center}
\vspace{-0.6cm}
\end{table}
Laboratory x-ray diffraction work was done on well-characterized crystals with an Agilent-Technologies SuperNova diffractometer using Mo-K$_{\alpha}$ radiation and a cryojet HT for temperature control. Above $T_{CO}$ (at 350$\,$K) the crystal structure of LuFe$_{2}$O$_{4}$ was refined \cite{WINGX} in $R\overline{3}m$ symmetry, with similar results as \cite{ISOBE90} and low $R$-factor $R[F^{2}\!>\!4\sigma(F^{2})]\!=\!1.87\%$. As already reported in \cite{Angst1,YAMADA1,YAMADA2} by cooling through $T_{CO}$ strong diffuse scattering along $(\frac{1}{3}\frac{1}{3}\ell)$ splits into sharp CO superstructure reflections (Fig.~\ref{fig:ACMS}a/b), with a small incomensurability. Only samples showing the best magnetic behavior, corresponding to those studied in \cite{Angst1,JOOST,XU,XU2,d0det}, show these sharp superstructure peaks already at room temperature. For refinements the apparent small incommensuration away from $(\frac{1}{3}\frac{1}{3}\ell)$ and $(00\frac{3}{2})$ type reflections was neglected, because it most likely corresponds not to a ``truly incommensurate'' structure \cite{Kim2006}, but rather to a discommensuration from anti-phase-boundaries as previously proposed for LuFe$_{2}$O$_{4}$ \cite{YAMADA2,Angst1} and also observed in other CO oxides \cite{OXO2}. The superstructure reflections originate from three individual CO domains \cite{Angst1} corresponding to 120$^{\circ}$-twinning with $(\frac{1}{3}\frac{1}{3}\frac{3}{2})$ and symmetry-equivalent $(\frac{1}{3}\overline{\frac{2}{3}}\frac{3}{2})$ and $(\overline{\frac{2}{3}}\frac{1}{3}\frac{3}{2})$ propagation vectors, as illustrated in Fig.$\,$1 of \cite{XU2}.\par

From symmetry analysis in the hexagonal cell with $(\frac{1}{3}\frac{1}{3}\frac{3}{2})$ propagation, two irreducible representations are allowed, both of which lower the space group to $C2/m$. These correspond to different origin positions (centers of inversion) of the monoclinic cell. In one case, it is located at the Lu positions between the bilayers, this structure corresponds to antiferroelectrically (AFE) stacked polar bilayers, as proposed in \cite{Angst1}. For the other case, the inversion center is located between the two Fe-layers of a bilayer, corresponding to (non-polar) bilayers with a net charge. This latter case was appraised as unlikely due to the necessity of inter-bilayer charge transfer \cite{Angst1}. However, only a full structural refinement can decide which representation is actually realized. A consequence from the domain structure is that some reflections totally overlap in reciprocal space, while others are difficult to separate, making a reliable refinement difficult.\par

Therefore, a quantity of small crystals, obtained from one crushed sample from the same batch as in \cite{d0det,JOOST,XU,XU2,Angst1} showing the best magnetic behavior \cite{JOOST}, was screened for their domain populations. In all experiments, the three domains were readily identified by the diffractometer software as twinned monoclinic cells with $C2/m$ symmetry. Most crystals show near-equilibrium populations (e.g. crystal 2 in Fig.~\ref{fig:ACMS}b), but some are close to a single-domain state (ratios of 0.03:0.09:1 for crystal 1), alleviating the structural refinement. On this crystal we collected 8556 reflections (1285 unique); all intensities were corrected by numerical absorption correction using indexed crystal faces.\par

A refinement in the structure model with the center of inversion located in the Lu-layers, corresponding to the representation with AFE stacked bilayers \cite{Angst1}, led to very anisotropic displacement parameters for Lu along the $c_{\mathrm{Hex}}$-direction. This is very unlikely for the heavy Lu ions. A relatively poor agreement was achieved: $R[F^{2}\!>\!4\sigma(F^{2})]\!\sim\!15\%$.\par

For refinements corresponding to the second representation with the center of inversion between the bilayers, a much better $R[F^{2}\!>\!4\sigma(F^{2})]\!=\!5.96\%$ is achieved. Additional refinements in lower-symmetry space groups, e.g. $Cm$, allowing for both CO configurations, reproduces a structure very close to this second, with only marginally improved $R$-values. This makes a lower symmetry than $C2/m$ very unlikely. The $C2/m$ structural solution is presented in Fig.~\ref{fig:Phaselow}a \cite{ICSD}. At 210$\,$K a Lu distortion along $c_{\mathrm{Hex}}$ with an amplitude of $\sim$0.14$\,$\r{A} (see Fig.~\ref{fig:Phaselow}b) is clearly connected to the Fe$^{2+/3+}$ CO involving O shifts on the Fe-O-Lu path, explaining the poor refinement with large anisotropic displacement parameters for Lu on high-symmetry sites (also visible as precursor effect in the hexagonal solution above $T_{CO}$ \cite{ISOBE90,ICSD}). For different Fe-sites strong deviations for the positions of surrounding O-atoms with respect to the high-$T$ structure are visible in Fig.~\ref{fig:Phaselow}c/d, indicating a separation into two Fe-valence states according to the average Fe-O bondlengths (Tab.~\ref{tab:Valences}). For Fe$^{2+}$ and Fe$^{3+}$ the average Fe-O bond length in a trigonal bipyramid coordination should be 2.09$\,$\AA$\,$ and 1.98$\,$$\mathrm{\AA}$, respectively \cite{SHANNON}.\par

\begin{figure}
\includegraphics[width=0.99 \linewidth]{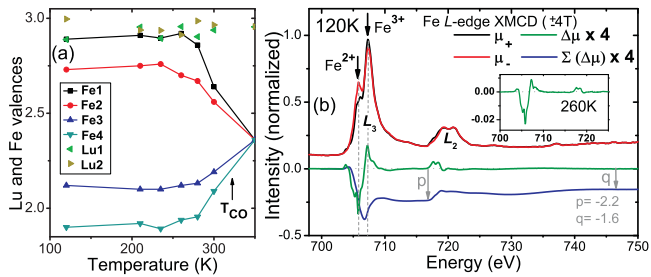}
\caption{\label{fig:BVS} (Color online) (a) Fe and Lu valence states for different sites from the bond-valence-sum method. (b) XMCD spectra across the Fe $L_{2/3}$-edge at 120$\,$K and 260$\,$K (inset). The XAS spectra with $H$ parallel and antiparallel to the incoming beam by changed photon polarization were averaged by subtracting them from each other \cite{Supple}.}
\end{figure}

To determine the valence $V$ from different cation sites a BVS analysis \cite{BONDVALENCE} was performed: $V=\sum_{i} \mathrm{exp}[(d_{0}-d_{i})/0.37]$. Here, $d_{i}$ are the experimental bond lengths to the surrounding ions and $d_{0}$ is a tabulated empirical value characteristic for the cation-anion pair \cite{BONDVALENCE}. The Lu-valences from BVS calculations are very close to $3+$ for all temperatures and sites. The results for the Fe-sites are shown in Tab.~\ref{tab:Valences} and illustrated by different shadings for different Fe sites in Fig.~\ref{fig:Phaselow} for the CO phase. The temperature dependence of the BVS (Fig.~\ref{fig:BVS}a) indicates below $T_{CO}$ an increasing valence separation upon cooling, with a plateau reached below 260$\,$K. At $T_{LT}\sim$170$\,$K there is a magneto-structural phase transition with a small splitting of structural reflections \cite{XU}. For the data at 120$\,$K an abnormal increase of the $c_{\mathrm{Hex}}$, calculated from the monoclinic lattice, is observed. However, in the refinements only very subtle changes of atom positions are achieved, not affecting the CO configuration (120$\,$K in Fig.~\ref{fig:BVS}) from BVS calculations, or the $C2/m$ symmetry.\par

The Fe valence separation on majority sites tends to be smaller than for the minority sites. The average valence of all Fe-sites from BVS is $\sim$2.4, the same as the Fe-BVS above $T_{CO}$, suggesting non-perfect ionicity of the bonds to the O despite of the large valence separation. The latter is also supported by a recent resonant x-ray diffraction study \cite{MULDERS}, in which a full 2+/3+ valence separation was deduced from the chemical shifts of the Fe $K$-edge. The valence separation deduced from BVS analysis is considerably larger than that for other CO Fe-oxides, except for Fe$_{2}$OBO$_{3}$ \cite{OXO}. \par

Is the new CO arrangement presented here consistent with the magnetic structure presented in \cite{JOOST}? Are the magnetic structures pre-determined by the CO? X-ray magnetic circular dichroism (XMCD) at the Fe $L$-edges is the ideal tool to address these questions. Two previous XMCD studies on LuFe$_{2}$O$_{4}$ were reported \cite{XMCD,XMCD2,XMCD3}, but both were performed on samples for which no long-range charge and magnetic order has been demonstrated. To test whether the strong spin-charge coupling deduced in \cite{XMCD,XMCD2} also applies to samples exhibiting long-range spin- and CO, we performed XMCD at the beamline 4-ID-C of the Advanced Photon Source (APS). We used magnetic fields up to 4$\,$T $||c_{\,\mathrm{Hex}}$ and the incoming beam and the total electron yield (TEY) as x-ray absorption spectra (XAS). The total fluorescence yield signal (30$^{\circ}$ between $k_{i}$ and $c_{\,\mathrm{Hex}}$) is dominated by re-absorption, but confirms the bulk nature of our XMCD \cite{Supple}. The XMCD signal was then calculated from the difference between the XAS (from TEY) for positive and negative circular polarization ($\mu_{+}$ and $\mu_{-}$), with no non-magnetic XMCD contributions \cite{Supple}. To see if there is any change in the CO configuration or structure between the two magnetic phases, we have done additional high-resolution x-ray diffraction at the beamline 6-ID-D (APS) above $T_{LT}$. The diffraction data in $H||c_{\,\mathrm{Hex}}$ up to 2.5$\,$T, show neither a change in the CO configuration nor a structural transition.\par

In the high-field ferrimagnetic phase \cite{JOOST}, the shape of the XMCD spectra $\Delta\mu$ (Fig.~\ref{fig:BVS}b) is similar to the ones shown in \cite{XMCD,XMCD2}. With the sum rule \cite{THOLE,CHEN} we could extract from $\Sigma(\Delta\mu$) a similar orbital-to-spin moment ratio of $\sim$0.3 corresponding to an orbital magnetic moment of $\sim$0.7$\mathrm{\mu_{B}}/$f.u., as previously reported \cite{THOLE,CHEN}. This observed unquenched orbital moment excludes the possibility of Fe$^{2+}$ orbital order for the ferrimagnetic phase; orbital order would imply a lifting of the two-fold-degeneracy of the lowest crystal-field doublet, which is occupied by a minority spin. However, this degeneracy is necessary for an orbital magnetic moment \cite{OOargument}. For the antiferro and paramagnetic phase it can also be excluded, indirectly, because there is no structural component in the transitions involved. Due to the structural transition at $T_{LT}$ \cite{XU} we can not exclude long-range orbital order in this low temperature phase, which could be consistent with the observed lattice parameter changes. However, a detailed discussion is beyond the scope of the present study.\par

Two prominent peaks in the $L_{3}$ region of the XAS are readily identified as the chemically shifted Fe$^{2+}$ and Fe$^{3+}$ white lines \cite{XMCD,XMCD2}. In the XMCD spectra the large downward-peak at the Fe$^{2+}$ position and smaller upward-peak at the Fe$^{3+}$ position, directly imply that the net moment of Fe$^{2+}$ is in field-direction and a smaller net moment of the Fe$^{3+}$ sites points opposite to the field. For the local Fe$^{2+/3+}$ spin configurations the model of \cite{XMCD,XMCD2}, extracted from a similar XMCD shape, together with the here presented CO is consistent with the ferrimagnetic spinstructure of \cite{JOOST}.

More important, the above implications of the XMCD signal, combined with the ferrimagnetic model \cite{JOOST}, verify the novel CO configuration. Given the absence of partial disorder only three valence specific local spin configurations are possible, of which only one is consistent with the overall magnetic saturation moment of $\sim$3$\,\mu_{\mathrm{B}}/\mathrm{f.u.}$ \cite{Supple}: All Fe$^{2+}$ as well as $\frac{1}{3}$ of the Fe$^{3+}$ spins are aligned in $H$-direction, $\frac{2}{3}$ of the Fe$^{3+}$ spins point opposite to $H$, the same model as proposed in \cite{XMCD,XMCD2}. Combining this local spin-charge configuration with the ferrimagnetic spin order \cite{JOOST} directly excludes any (anti)ferroelectric model preserving mirror symmetry. Ignoring mirror symmetry, 28 configurations are possible \cite{ENDNOTE}, of which however only the one presented in Fig.~\ref{fig:Phaselow}a fits to the right intensity distribution along the $(0\,2\,\ell)_{\mathrm{Mon}}$ line (Fig.~\ref{fig:ACMS}b); this is also the only one of the 28 preserving mirror symmetry.\par

Furthermore, as discussed in \cite{JOOST} the refinement of spin structures can be improved by introducing different magnetic moments for Fe$^{2+}$ and Fe$^{3+}$ according to the charged-bilayer CO model, which is not the case for any CO with polar bilayers. This also supports the above analysis, though by itself the weight of this evidence is reduced by a similar improvement of the refinement regarding a possible magnetic contamination. \par

Thus, structure refinement, XMCD, and magnetic contrast, all clearly identify the CO configuration with charged bilayers (Fig.~\ref{fig:Phaselow}a) as the correct one. This charge pattern is very surprising, because it requires inter-bilayer charge transfer. For this reason it was considered before only in \cite{Angst1}, where it was mentioned as symmetry-allowed but excluded as physically unlikely. Understanding the origin of this long range ($\sim6\,\mathrm{\AA}$) charge transfer calls for further theoretical work.\par

Importantly, this new CO structure does not contain polar bilayers, in contrast to what was previously proposed (e.g.\ \cite{Ikeda,Angst1}), casting doubt on the ``ferroelectricity from charge ordering" scenario. How general is our result, given the significant reported (see, e.g., \cite{Michiuchi2009}) sample-to-sample variations? Clearly, the structure refinement can be expected to be representative for all samples where $(\frac{1}{3},\frac{1}{3},\mathrm{halfinteger})$-reflections are observed as main CO order parameter (e.g.\ \cite{Ikeda,Mulders2,Angst1,WEN10,XU,XU2,MULDERS,d0det,JOOST,YAMADA1,YAMADA2,JOOST}), the similarity of observed XMCD spectra with \cite{XMCD,XMCD2} even suggests that the same basic CO configuration also applies to samples without long-range CO (e.g.\ \cite{WU}). In particular, our refinement should be valid for the samples on which pyroelectric current measurements have been reported \cite{Ikeda}.\par

To explain the pyroelectric current measurements, some of us proposed \cite{Angst1} that a ferroelectric CO might be stabilized by an electric field, though such a scenario seems less likely when charged bilayers have to be polarized. Indeed, a CO remaining completely robust in electric fields has been reported by Wen \textit{et al.} \cite{WEN10}, based on neutron diffraction. We have confirmed this as also valid for our samples by additional x-ray diffraction measurements in electric fields up to 15$\,$kV$\,$cm$^{-1}$ at APS 6-ID-D and 6-ID-B, and thus conclude that a ferroelectric CO cannot be stabilized by electric fields.\par

The relatively low resistivity around $T_{CO}$ \cite{WEN10,Zeng2008,Li2008,Fisher2011,REN} could provide an alternative explanation for the pyroelectric current measurements of \cite{Ikeda}, because in the presence of residual conductivity non-ferroelectrics can exhibit currents strongly resembling ferroelectric depolarization currents, due to space-charge effects \cite{Maglione2008}. The also observed \cite{Ikeda,SUBMANIAN} giant dielectric constants could be attributed to interface effects \cite{REN}. All reported macroscopic indications of ferroelectric behavior in LuFe$_2$O$_4$ are therefore most likely due to extrinsic effects. \par

Returning to XMCD, the analysis not only shows the consistency of the new CO and the spin order, but also implies a strict coupling of these orders. Interestingly, XMCD spectra taken above $T_{N}$ in $\pm4\,$T (Fig.~\ref{fig:BVS}b inset) have a small amplitude, but indicate the same Fe$^{2+/3+}$ spin configuration as in the ferrimagnetic phase. This is consistent with the conclusion for $H\!=\!0$ of randomly-stacked bilayers that are still individually magnetically ordered based on diffuse magnetic scattering \cite{JOOST}: partial polarization by a magnetic field is then expected to lead to the same relative net moments on Fe$^{2+}$ and Fe$^{3+}$, provided the spin-charge coupling remains. This signifies still ordered Fe-bilayers in the paramagnetic phase with strictly coupled charge- and spin order persisting well above T$_{N}$, from susceptibility also likely above $T_{CO}$. This suggests at high $T$ short-range precursors with already coupled local spin- and CO. This coupling already above $T_{CO}$ is most likely the origin of the magnetic-field control of charge structures reported in \cite{WEN09}.\par

In conclusion, crystal structure refinements of charge ordered LuFe$_{2}$O$_{4}$ show that the Fe/O bilayers are charged rather than polar. This is further supported by an analysis of XMCD data, which also indicates a strict spin-charge coupling extending to the fluctuations-regime above the ordering temperature. The non-polar CO, which is not affected by electric fields, precludes CO-based ferroelectricity in LuFe$_{2}$O$_{4}$. Hence, a clear example of an oxide material with ferroelectricity originating from CO has yet to be identified. 

We gratefully acknowledge D. Robinson for help with collecting data. Support from the initiative and networking fund of the Helmholtz Association of German Research Centers by funding the Helmholz-University Young Investigator Group ``Complex Ordering Phenomena in Multifunctional Oxides'' is gratefully acknowledged. Use of the Advance Photon Source was supported by the U.S. Department of Energy, Office of Science, Office of Basic Energy Sciences, under Contract DE-AC02-06CH11357. MA thanks D. Mandrus, B.C. Sales, W. Tian and R. Jin for their assistance in crystal growth, also supported by the US DOE.


\begin{thebibliography}{99}
\bibitem{ICSD}
The structures refined are deposited in the Inorganic Crystal Structure Database with access numbers:
423570, 423569 (350$\,K$ and 210$\,$K).
\bibitem{ENDNOTE}
All four $\downarrow$-sites must be Fe$^{3+}$, and the number of possibilities to distribute two remaining Fe$^{3+}$ among the 8 $\uparrow$-sites is ${8 \choose 2}=28$.
\end{thebibliography}
\end{document}